\documentclass[]{article}

\usepackage{epstopdf}% To incorporate .eps illustrations using PDFLaTeX, etc.
\usepackage[caption=false]{subfig}% Support for small, `sub' figures and tables

\usepackage{biblatex}
\usepackage[]{url}
\usepackage{amsmath}
\usepackage{graphicx}

\title{On the use of chaotic dynamics for mobile network design and analysis: towards a trace data generator}

\author{
Martin Rosalie\textsuperscript{a, b} and Serge Chaumette\textsuperscript{c} \\
}

\begin{document}

\maketitle

\noindent\textsuperscript{a}Univ. Perpignan Via Domitia, Laboratoire Génome et Développement des Plantes, UMR 5096 Perpignan, F-66860, France

\noindent\textsuperscript{b}Centre National pour la Recherche Scientifique, Laboratoire Génome et Développement des Plantes, UMR 5096 Perpignan, F-66860, France

\noindent\textsuperscript{c}Univ. Bordeaux, LaBRI, UMR 5800 Talence, F-33400, France

\noindent\textbf{Correponding author:} M. Rosalie. Email: \url{martin.rosalie@univ-perp.fr}

\begin{abstract}
    With the constant increase of the number of autonomous vehicles and
    connected objects, tools to understand and reproduce their mobility models
    are required. We focus on chaotic dynamics and review their applications
    in the design of mobility models. We also provide a review of the
    nonlinear tools used to characterize mobility models, as it can be found
    in the literature.  Finally, we propose a method to generate traces for a
    given scenario involving moving people, using tools from the nonlinear
    analysis domain usually dedicated to topological analysis of chaotic
    attractors.
\end{abstract}

\noindent\textbf{Keywords}
 Mobile networks; Mobility models; Chaotic dynamics;
Poincaré section; First return map

\section{Introduction}

The number of applications that use autonomous devices, for instance robots or
Unmanned Aerial Vehicles (UAVs), increases nowadays. In this context,
defining and analysing their mobility is particularly important. A mobility
model describes the behaviour of an entity considering its capacities, possible
moves and speed. The mobility models are described either analytically at the
individual level, or by the interactions between the parts of the system
(between UAVs, UAVs and planes, UAVs and points to survey, etc.). The
resulting behaviours described with these simple rules can induce the
emergence of a global intelligent behaviour. Inversely, from the resulting
behaviour of such a swarm, these initial simple rules are hard to discover. A
similar phenomenon occurs for chaotic dynamics where a chaotic process appears
to be random while it arises from a deterministic process. Therefore, the idea
is to study the connections between the two concepts.

Literature on \emph{chaotic dynamics} and \emph{nonlinear dynamics} has been
developed at the end of the century while the main concepts of this theory
come from the early nineties (see Fig.~1 of \cite{aguirre2009modeling}). Chaos
is observed, described and analysed in numerous domains, for instance:
electronic circuits \cite{matsumoto1985double}, chemical reactions
\cite{rossler1976equation}, laser behaviours \cite{meunier1992combined} or
biological models \cite{yao1990model}. The first example of the use of chaos
(from the Chua system \cite{chua1986double}) to design a mobility model has
been proposed in 1997 \cite{saiwaki1997automatic}. The authors proposed to use
the properties of chaotic dynamics to ensure a good coverage of an area.
Chaotic dynamics is defined as follows:   the solution of a deterministic
process is chaotic if it is sensitive to initial conditions, aperiodic and
globally time invariant. These properties induce that this chaotic solution
will be unpredictable   when considering a long term behaviour. Nowadays,
several tools improve the understanding of a chaotic behaviour, chaotic
mechanism and bifurcation diagrams and some authors use them to provide
mobility models.  Because ``\emph{little is known of the potential
relationship between swarm and chaotic systems}''
\cite{harvey2015application}, the main goal of this paper is to explore the
uses of chaotic dynamics in the domain of mobility models.  This can be done
either by designing chaotic deterministic models or by using tools coming from
nonlinear analysis.

In this paper we first present a review of mobile networks including chaotic
dynamics using chaotic maps or ordinary differential equations. We then review
the tools used to analyse these systems. In section 3, we propose a method to
generate traces using nonlinear analysis tools. This method is supported by
numerical simulations using the Lorenz system and permits to reproduce
congestion and distribution patterns.  We finally present the conclusion along
with our future work in section 4.

\section{Mobility models and chaotic dynamics}
\label{sec:state_of_the_art}

In this section, we present models from the literature including chaotic
dynamics and the tools used to analyse them. In these models, chaos is
obtained from well-known discrete or continuous systems. The chaotic variables
of these systems are used to design mobility models.  Sections \ref{ssec:map}
and \ref{ssec:ode} detail the two main approaches used to generate chaotic
dynamics for mobility models using: a chaotic  attractor from a discrete
system or a chaotic attractor solution of a continuous system.  Section
\ref{ssec:tools} is dedicated to the tools from nonlinear analysis and their
applications.  The last section (Sec.~\ref{ssec:CACOC}) details our previous
contributions in this domain in which we proposed to use periodic orbits of
chaotic dynamics to define mobility models of UAVs in order to enhance the
coverage of an area.

\subsection{Models using chaotic maps}
\label{ssec:map}

\emph{Chaotic maps} are nonlinear recurrent relations. The most used in the
literature is the \emph{logistic map} defined by $x_{n+1}=\alpha x_n(1-x_n)$.
Introduced by Verhulst \cite{verhulst1845recherches} it represents the growth
of population $x$ at each step where $\alpha$ is the growth rate. This map
contains only one nonlinear term and can exhibit both periodic and chaotic
dynamic.  When varying $\alpha$ from 2 to 2.4, its bifurcation diagram results
in a period doubling cascade: a classical route to chaos found in several
systems. For details on the logistic map, the reader is referred to
\cite{Boeing2016}.  Charrier \textit{et al.} \cite{charrier2007nonlinear}
propose to use the logistic map to reproduce the behaviour of flocks. In the
original paper \cite{reynolds1987flocks}, Reynolds introduced \textit{boids}
to reproduce the flocking behaviour with three rules (collision avoidance,
velocity matching and flock centring) that generates a force vector for each
agent in the swarm. In the Charrier \textit{et al.} model the synchronization
between agents is performed by the environment: the control parameter of the
logistic map is updated depending on the neighbourhood of each agent. This
model does not use the standard rules of flocking but reproduces their
behaviour. The authors use a \textit{bifurcation diagram} to emphasize the
convergence of the system and to show the global dynamics of their agents and
the transition to chaotic dynamics.

A chaotic map (the standard map \cite{wallis1984}) can also be used to produce
chaotic motion for a robot \cite{martins2007trajectory}. From such a
two-dimensional map, a planning is assigned to the robot using two coordinate
points obtained from the chaotic map. Then a robot visits these points in
their order of appearance. As the purpose of the robot is to cover a square
surface, the authors evaluate their system using coverage rate.  Curia
\textit{et al.} \cite{curiac2014chaotic} proposed to use another map, the
Hénon map, to   move a mobile robot with unpredictable trajectories.  Their
mobility model combines a guiding line that the robot follows with a chaotic
motion obtained from the map. While the robot follows the guideline, the
chaotic motion controls the evolution of the robot around this line. Here
again, the authors use a bifurcation diagram to  underline the chaotic
properties of their system. Even if their system is made of six equations
(including the Hénon map), they prove that it has a chaotic solution when
$a=1.4$ as it is the case for the Hénon map. The additional equations
dedicated to the movement around the line do not influence the chaotic
dynamics.

\subsection{Models using ordinary differential equations systems}
\label{ssec:ode}

In this section, we  present models that use a set of ordinary differential
equations as a source of chaotic dynamics.  We are now considering chaotic
continuous solutions instead of discrete solution from a map. However, there
is a way to discretely represent these continuous chaotic dynamics. These
solutions are embedded in the \textit{phase space}, for instance the Lorenz
attractor \cite{lorenz1963deterministic} is a famous attractor in a three-dimensional
phase space. Introduced by H. Poincaré
\cite{poincare1893methodes}; the \textit{Poincaré section} is a transversal
surface of the flow  that provides a discrete description of the chaotic
dynamics obtained from continuous systems.  The original idea is to consider
only the discrete points when the flow crosses a Poincaré section instead of
the whole trajectory in the phase space.  The discrete sequence of points
contains a synthesis of the dynamical properties of the attractors. Most of
the references presented below use the chaotic dynamics from the chaotic
attractor using a Poincaré section.

There are many examples of such systems where nonlinearity induces chaotic
behaviour. Among the most studied one: Lorenz, Rössler, Chua systems; there
are several articles about robots, the mobility models of which use a set of
ordinary differential equations.  For instance, Nakamura \& Sekiguchi
\cite{nakamura2001chaotic} use the Arnold equations to model the behaviour
of a mobile robot with chaotic motion. They prove that the coverage of their
chaotic robot is better than the coverage of a robot using a random walk.
Further to this work, Bae \textit{et al.} \cite{bae2004target} proposed a
``chaotic UAV'' with the Chua system, Arnold equations and Van Der Pol
equations. They also introduce an obstacle avoidance method without decreasing
the performance of the model in terms of coverage.

Fallahi \& Leung \cite{fallahi2010cooperative} proposed a cooperative set of
four mobile robots synchronized using Chen \cite{chen1999yet} and Lorenz
\cite{lorenz1963deterministic} systems. They used one of the variables of
these systems to define the movements of their robots.  In their system, one
of the robots is the master, and the others are synchronized with   it. This
system is efficient compared to unsynchronized robots or random walks in terms
of coverage rate and travelled distances. Similarly, Mukhopadhyay \& Leung
\cite{mukhopadhyay2013cluster} present synchronized robots with chaotic path
planners.  They add a \textit{symbolic dynamic} description of their robots.
The symbolic dynamic is a nonlinear analysis tool used to describe chaotic
dynamics. Its purpose is to label a discrete trajectory obtained from a
Poincaré section. Thus, it gives a symbol according to the dynamical aspect
of the solution depending on the topological period of the system. At the end,
the solution is no longer a variable but a sequence of symbols indicating
dynamical aspects of the system studied.  The authors use this sequence of
symbols to evaluate the synchronization rate between their robots.  We also
would like to mention the work of Bezzo \textit{et al.}
\cite{bezzo2014decntralized} where synchronization of chaos is used to detect
changes in the topology of a mobile robotic network. The authors study the
motion of mobile agents through an unknown environment with obstacles (see
also \cite{Sorrentino2008,Sorrentino2009} for details on this method).

Volos and co-workers \cite{volos2012chaotic} proposed another use of chaos
which consists in designing a path planning generator for autonomous mobile
robots. From the double scroll chaotic circuit (Chua system
\cite{chua1986double}), they obtained a chaotic true random bit generator.
They use it to define the path planning of the robots  \textit{i.e.}, the list
and the order of  the points the robots have to visit. The efficient coverage
rate and the unpredictability of the robot trajectories are the main
characteristics of this system.  This system is similar to the system using
chaotic map \cite{martins2007trajectory}. Comparing their coverage rate in
terms of the number of planned points, the system designed by Volos and
co-workers \cite{volos2012chaotic} is ten times better.  From the same
authors, similar results are obtained when the Arnold map is used to generate
waypoints for path planning \cite{Curiac2015}. Finally, we would like to point
out a recent work done by Pimentel-Romero and co-workers
\cite{pimentel2017chaotic} using Poincaré sections of chaotic attractors as
threshold to generate random numbers. They conclude that these particular
Random Number Generators (RNGs) using chaotic dynamics are efficient  for
generating random paths for autonomous mobile robots.

Another way to include chaos  to support mobile robot mobility is presented
by Rosyid \textit{et al.} \cite{rosyid2012performance}. Their method does not
use any well-known  Ordinary Differential Equations (ODE) system to drive the
robots. Robots communicate using sound that all can hear.  Each robot moves in
a direction depending on the ``total amount'' of sound received. This is
modelled by an ODE that details how this system is synchronized because their
relative positions influence the sound emitted  and received. The authors use
the coverage rate to compare their system to the previously presented methods
for robots driven with Lorenz or Arnold equations to prove that their model
has better performance. They also compute the \emph{Largest Lyapunov Exponent}
(LLE) to ensure that chaos occurs in their system.   This value is a measure
of the separation rate of two infinitely initially closed trajectories
\cite{wolf1985determining,ott2002chaos}. The LLE refers to the predictability
of a system and is commonly used as an indicator of a chaotic behaviour. This
is a metric approach that does not permit to distinguish chaotic solutions
that have distinct structures in the state space. The reader is referred to
\cite{mindlin1990classification,gilmore1998topological,byrne2004distinguishing}
for details about chaotic mechanisms (\textit{e.g.} folding mechanism or
tearing mechanism).  The 0--1 test can also be used as an indicator to
distinguish chaotic dynamics from periodic one (the reader is referred to
\cite{gottwald2004new} for details).  We also mention that diffusion
coefficient can be computed for first return maps to measure the difference
between the dynamics when a parameter is varied
\cite{grossmann1982,klages1995}.

We recently proposed mobility models
\cite{rosalie2016random,rosalie2017coverage,rosalie2018chaos} using chaotic
behaviour based on the Rössler system using Poincaré section and periodic
orbits. In the next section we will present nonlinear tools used to analyse
mobility models. Then, in section~\ref{ssec:CACOC}, we present our mobility
models and the nonlinear tools used to build and analyse them.

\subsection{Nonlinear analysis tools}
\label{ssec:tools}

We  gave above examples of mobility models built from chaotic dynamics.  Some
of them have been analysed with tools coming from the domain of nonlinear
analysis. In this section we present  studies carried out on mobility models
using these nonlinear tools to understand and describe their behaviour.

In 2014, Timme \& Casadiego wrote an article entitled ``\textit{Revealing
networks from dynamics}'' \cite{timme2014revealing}. This paper gives an
overview of approaches considering collective nonlinear network dynamics, but
few details are given about tools from the domain of nonlinear analysis that
could be used when chaotic systems are identified.  These tools are well
introduced by Qu \textit{et al.} \cite{qu2006emergence} in their paper about
emergence in swarming  systems. The authors detail the emergence phenomenon
and the following tools: Lyapunov Exponents, Attractor, Recurrence plot,
Poincaré section. They also extract \textit{periodic orbits} from a
Poincaré section.  The ``periodic'' term of periodic orbits refers to the
state space and not to the time space  (topological period). Periodic orbits
are time invariant while the system evolves in a chaotic  state (from initial
condition, the solution evolves and successively visits the unstable
periodic orbits). From a chaotic time series and with a Poincaré section,
orbits can be extracted, and this acquisition is a preliminary step of the
\textit{topological characterization} \cite{gilmore1998topological}. For
dissipative systems, the purpose of this method is to obtain the structure of
the chaotic mechanism from a topological invariant (the linking number)
computed between periodic orbits (the reader is referred to
\cite{gilmore2002topology} for details).

Hazan \textit{et al.} \cite{hazan2006topological} opt for this approach
because the aim of their work is to classify the behaviour of robots using
periodic orbits. The authors explain that their method is not based on a
metric (for instance the LLE) because it is ``highly sensitive to
perturbations such as noise contamination'' \cite{hazan2006topological}. They
use the topological characterization tool after building an embedding (they
reconstruct a phase space from one variable) of the behaviour from the
$x$--axis motion of the robot. They build a Poincar\'e section in order to
extract periodic orbits and then describe the behaviour of the robots using
the linking numbers between the orbits.

Das and co-authors \cite{das2014stability,das2015chaotic} propose to use
Lyapunov exponents to distinguish transitions in a multi-agent swarm system.
The system is designed to solve an optimization problem where the agents have
to reach a particular point. Computing Lyapunov exponents enables the authors
to find the range of parameters where chaos occurs and where the system is no
longer periodic. The authors propose an application  of their system: each
robot is an automatic fire extinguisher, and they have to reach a burning
place.  ``\textit{In any swarming dynamics, emergence of chaos is a very
important situation to be dealt with}'' \cite{das2015chaotic} and this is
illustrated by the work of Wu and co-workers \cite{wu2011analysis}. In the
latter article, the authors use Lyapunov exponents to analyse their swarming
system and concluded that the chaos in swarm model becomes weaker while the
emergence becomes stronger. The Lyapunov exponents are also used to analyse a
swarm model of Self Propelled Particles (SPP) by Shiraishi and Aizawa
\cite{shiraishi2014lyapunov,shiraishi2015collective}. This tool permits to
understand  the relations between the behaviour of the system and the number of
agents: ``\textit{the Lyapunov exponents reflect the biological sensitivity
hidden behind the motion of swarm}'' \cite{shiraishi2014lyapunov}.

Chaos in neuronal network is also studied with Lyapunov exponents
\cite{eser2014nonlinear} or more recently, using reconstructed attractors with
time-delay coordinates and with a Poincar\'e section
\cite{likhoshvai2015alternative}. These latest tools are robust to well define
the chaotic mechanism because the first return map to the Poincar\'e section
with unimodal structure indicates that there is a stretching and folding
mechanism: it is a signature of the classical ``horseshoe'' mechanism (also
known as \textit{folding mechanism}).  This type of chaotic mechanism is also
present in the work of Sato and co-workers
\cite{sato2003coupled,sato2005stability} as illustrated by Fig.~6 of
\cite{sato2005stability}. They work on a multiagent system using reinforcement
learning that is modelled with coupled differential equations. This is applied
to game theory: Matching Pennies and Rock-Scissors-Paper games. The stretching
and folding mechanisms describe the effect of mutual adaptation and memory
loss with non-transitive structure for their system. This leads to Hamiltonian
chaos if there is no memory loss and to a dissipative system where there is
memory loss. The dissipative system exhibits limit cycles, intermittency and
deterministic chaos. To study their system, they employ Lyapunov exponents,
Poincar\'e section, bifurcation diagram and extract periodic orbits.  These
tools are also used to study languages and learning mechanisms where chaotic
dynamics appears \cite{mitchener2004chaos}.

\subsection{Mobility models based on periodic orbits}
\label{ssec:CACOC}

We recently proposed  mobility models using chaotic behaviour based on the
R\"ossler system
\cite{rosalie2016random,rosalie2017coverage,rosalie2018chaos}. These mobility
models permit to enhance the coverage of an area compared to random mobility
models.  We used the \textit{first return map} from a Poincar\'e section of a
chaotic attractor solution of the R\"ossler system and considered the periodic
orbits to build  efficient mobility models in terms of coverage rate.  The
Rössler system \cite{rossler1976equation} is given by the equations
\begin{equation}
    \left\{\begin{array}{rl}
  \dot{x} &= -y -z \\
  \dot{y} &= x +ay \\
  \dot{z} &= b+z(x-c)
\end{array}\right.
\label{eq:rossler}
\end{equation}
and its Poincaré section is defined as follows:
\begin{equation}
    P = \{ (y_n, z_n) | x_n = 0, \dot{x}_n > 0 \}
    \label{eq:rossler_section}
\end{equation}

\begin{figure}[htpb]
    \centering
    \includegraphics[width = .65\textwidth]{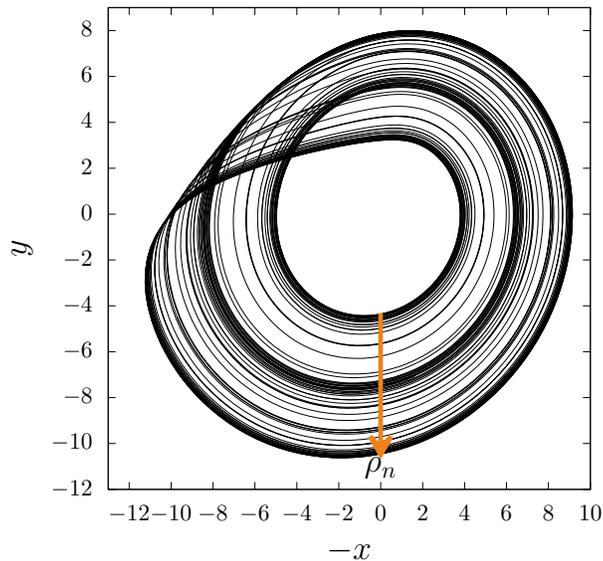}
    \caption{
        Chaotic attractor solution of the Rössler system \eqref{eq:rossler}
        (values of parameters $a=0.1775$, $b=0.215$ and $c=5.995$)
        with the Poincaré section \eqref{eq:rossler_section} represented by
        an arrow. % The flow evolve clockwise in this projection.
    }
    \label{fig:rossler}
\end{figure}

\begin{figure}[bp]
    \centering
    \includegraphics[width=.45\textwidth]{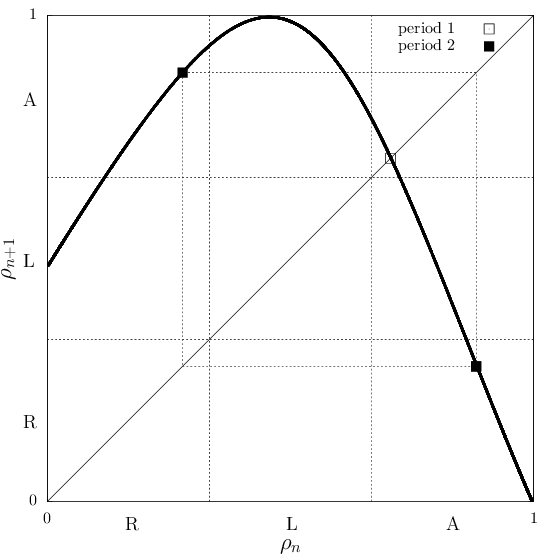}
    \caption{
        First return map to the Poincaré section of the R\"ossler attractor
        (Fig.~\ref{fig:rossler}).  This map is partitioned in three parts
          that give the UAV directions: L (left), A (ahead) and R (right).
        Orbits of period 1 and 2 illustrate patterns (AAAAA\dots) and
        (ARARA\dots), respectively straight lines and large turns. These
        patterns are efficient to cover an area with UAVs
        \cite{rosalie2018chaos}.
    }
    \label{fig:map}
\end{figure}

These tools are used to obtain the topological structure of the Rössler
attractor \cite{rosalie2014toward} where $\rho_n \in [0;1]$ is the normalized
value of $y_n$ in the Poincaré section (Fig.~\ref{fig:map}).  We introduce a
new concept to provide trajectories for UAVs: the dynamics of the first return
map enable us to obtain a local direction. The first return map is a step-by-step
process used to update the direction of the UAVs based on a three symbols
dynamic (L for left, R for right and A for ahead).  The periodic orbits of an
attractor are considered as its skeleton because they structure the dynamics
of the system. From the first return map, we extract the periodic orbits of
attractors to obtain these recurrent points often visited with the same order.
Thus, our UAVs  can follow these specific patterns that  allow them to explore
a wide area.  We obtained straight lines and wide turns with respectively
period one orbit (AAAA \dots~for ahead, ahead, ahead, \dots) and period two
orbit (ARARAR \dots~for ahead, right, ahead, \dots) (Fig.~\ref{fig:map}).  The
period one orbit leads to a straight forward line to enable exploration while
the period two orbit  enables the UAV to make large right turns to change  its
direction.  The reader is referred to \cite{rosalie2018chaos} for details
about the periodic orbits of attractors for mobility models of UAVs.  The
increase in performance provided by these mobility models using first return
maps indicates that they deserve further investigation. In the next section,
we propose a method to generate traces for mobile agents using this specific
tool from nonlinear analysis.  We would like to answer the following question:
from a global point of view, can first return maps be useful to produce and/or
analyse traces of UAVs?

\section{Traces generation from multi-components Poincaré section}
\label{sec:position_paper_on_trace_generation}

In this section we first present the concept of \textit{partial first return
map} as a tool to describe data with an unknown part. Conversely to the
classical first return map, this partial first return map allows input and
output which is a mandatory property to provide data traces.  The main purpose
of this article is to analyse and generate traces of agents in an open
environment which means that the agents can be added or removed in the model.
Then, section \ref{ssec:scenario} provides a scenario where a partial first
return map can describe the behaviour of agents.  In section
\ref{ssec:experimental}  we present a proof of concept supported by numerical
experiments using a theoretical dynamical system: the Lorenz system
\cite{lorenz1963deterministic}.

\subsection{Concept}

The purpose of this section is to propose a methodology to generate traces of
mobile entities named \emph{agents} (robots, persons, UAVs, \dots) using the
most accurate tools related to the chaotic behaviour. We consider that the
agents move in a well-defined area (the environment).  The agents  can enter,
move and exit  this area after a while. We suppose that the behaviour of the
agents is mainly induced by constraints of the environment that force them to
follow certain paths . As a consequence, we can make an analogy between the
environment and the phase space of a deterministic dynamical system. We do
consider that these paths are similar to unstable periodic orbits of  an
attractor in  a phase space that agents might follow.

The classical method applied in  this case is to reconstruct the whole phase
space from one measured variable (for instance one coordinate of the
position). Contrary to Hazan \textit{et al.} \cite{hazan2006topological}, we
do not consider that the traces contain the whole dynamical properties, but
only a portion of it because of the  capability of the considered robot to
enter and exit the area.  They use the standard way to reconstruct the phase
space. For the same reason, we are not able to use even better global
modelling methodology for time series data \cite{mangiarotti2014chaotic}.

Based on our experiment in mobility models design
\cite{rosalie2016random,rosalie2017coverage,rosalie2018chaos}, we can say that
an approach using the orbits of an attractor as guidelines is very efficient
in terms of coverage of an area.  This efficiency is due to the patterns
followed by the UAV during the exploration process. Such an approach including
patterns repetition can be applied to  the generation of traces of mobile
agents. Fig.~\ref{fig:map} is the first return map of a Poincar\'e section of
a R\"ossler attractor used in  several research projects
\cite{rosalie2016random,rosalie2017coverage,rosalie2018chaos}. This first
return map is an unimodal map made of an increasing branch and a decreasing
branch: this illustrates the folding mechanism (``horseshoe'' mechanism).
However, this first return map details the whole chaotic dynamics of a bounded
and globally time invariant process of a given chaotic attractor. Thus, the
periodic points of Fig.~\ref{fig:map} describe the entire periodic orbits and
considering the analogy with the mobility model, this prevents entrance or
exit of agents. As we are aiming to obtain a trace data generator, this global
perspective is a drawback to overcome, because the agents cannot be considered
as permanently evolving in a dedicated  well-defined environment.
Therefore, we propose a new method  using  partial data from a  chaotic
attractor.

\begin{figure}[htpb]
    \centering
    \includegraphics[width = .8\textwidth]{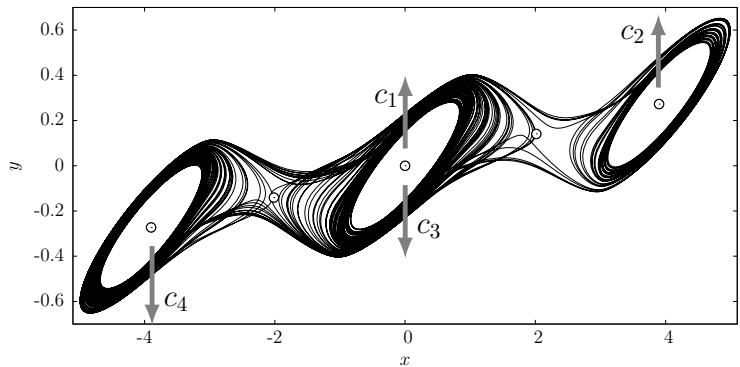}
    \caption{The multispiral attractor defined in \cite{alaoui1999differential} with the
    fixed points and the four components of the Poincar\'e section: $c_1$, $c_2$, $c_3$
    and $c_4$  \cite{rosalie2014toward}.
        This
    attractor is bounded by a genus--5 torus (the five aligned holes of the bounding torus are the fixed points and indicated
    with a dot in a circle $\bigodot$).}
    \label{fig:alaoui}
\end{figure}

For attractors bounded by high genus torus, the Poincar\'e section is made of
several components  that can be used to properly describe  chaotic dynamics
using symbolic dynamics \cite{rosalie2014toward}. For instance, the
multispiral chaotic attractor (Fig.~\ref{fig:alaoui}) introduced by
Aziz-Alaoui \cite{alaoui1999differential}  is bounded by a genus--5 torus. To
describe the dynamics in a discrete way, the Poincaré section has to detail
the transitions between the spirals. Consequently, the Poincaré section will be
made of several \textit{components}. These components  are chosen accordingly
to the bounding torus theory \cite{tsankov2003strange} using the fixed points.
For the multispiral attractor (Fig.~\ref{fig:alaoui}) this theory indicates
that four components  are required to build the Poincar\'e section.  The
Poincaré section is no longer one plane but a set of planes.  The flow crosses
these planes and from a continuous flow we obtain discrete values.  We used a
concatenation of these values  to build one variable representing the whole
Poincaré section: $\rho_n$. Each component   is represented by a range of
values in $\rho_n$: component $c_i$ for $\rho_n \in [i-1,i]$ (see
Fig.~\ref{fig:alaoui}).  Thus, $\rho_n$ is a variable between 0 and 4 that
describes the Poincaré section. The four components  synthesized in $\rho_n$
describe the entire dynamic of the system.

\begin{figure}[tbh]
    \centering
    \includegraphics[width = .7\textwidth]{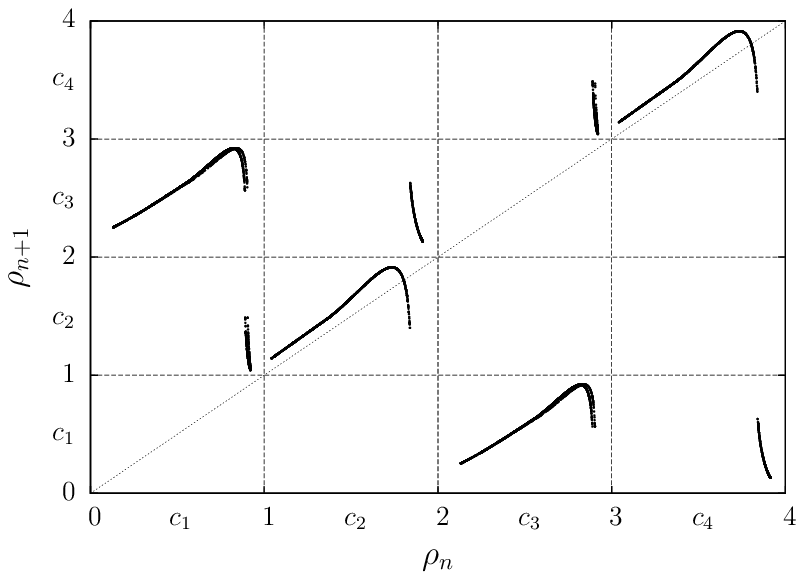}
    \caption{
        First return map of a Poincar\'e section describing a multispiral
        attractor (Fig.~\ref{fig:alaoui}) using a Poincar\'e section
        made of four components \cite{rosalie2014toward}.
    }
    \label{fig:4component}
\end{figure}

In a first return map to this Poincaré section,  orbits can be extracted, and
a symbolic dynamic can be assigned  as it has been done for the Rössler
attractor (See Fig.~\ref{fig:map} that details periodic points associated to
periodic orbits). Fig.~\ref{fig:4component} shows a first return map of a
Poincaré section made of four components for the multispiral attractor
(Fig.~\ref{fig:alaoui})  by plotting $\rho_{n+1}$ versus $\rho_n$. This first
return map contains both horseshoe mechanisms and tearing mechanisms. This
map is a discrete description of the flow and permits to obtain information
concerning possible transitions between components of the Poincaré section.
The reader is referred to \cite{rosalie2014toward} for details on the
methodology used to build  a Poincar\'e section with several components using
the properties of fixed points and the bounding torus theory.
One of the way to build a bounding torus is to place the hole where the fixed
point of the attractor are. For the multispiral attractor there are details
in \cite{rosalie2014toward} and for the Rössler system, there are details in
\cite{rosalie2016templates} including parameters variation ensuring the
robustness of the method.

In this multispiral attractor there are transitions from a component to the
same component (\textit{e.g.} $c_2$ to $c_2$) and transitions to another
component (\textit{e.g.} $c_2$ to $c_3$). The first return map
(Fig.~\ref{fig:4component}) details the chaotic mechanism occurring to perform
these transitions and describes the whole dynamics of the system.  The novelty
of our approach lies on the use of only a part of the data to introduce
chaotic mechanisms as  a model for traces generation.  We develop a new tool
to handle this unknown part of the data: the \textit{partial first return
map}.  A partial return map is an incomplete map with at least one component
with incoming flow, the \textit{initial components}, and at least one
component with outgoing flow, the \textit{final components}. The flow is split
in such a way that the partial first return map describes the transitions
between the initial components and the final components using
\textit{transitional components}. With this repartition of components, we
propose to follow a particular ordering to build the partial first return map:
the initial components, the transitional components and the final components.

Our new concept is to consider only a part of the trajectory  in the
environment.  Thus, a \textit{partial first return map} is only  an uncompleted
first return map with  transitions between a subset of components of  a
Poincar\'e section.  This permits to represent experimental data without
taking into account the rest of the trajectories. Consequently, this missing
or unknown data can be considered as input and output for traces model
indicating where agents enter or leave the environment. In the next section we
first present a scenario for our methodology and in section
\ref{ssec:experimental} we detail the method to build partial first return
map.

\subsection{Scenario}
\label{ssec:scenario}

\begin{figure}[htb]
    \centering
    \includegraphics[width=.35\textwidth]{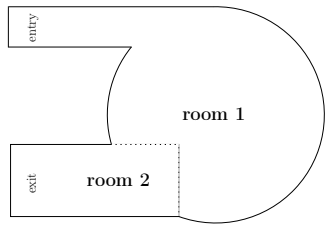}
    \caption{
        Scenario using an exhibition centre with two rooms. The shape is
        intentionally similar to the considered phase space
        (Fig.~\ref{fig:lorenz_A_attra_section}) to underline the asset of our
        method but this is not mandatory: only transitions between components
        are significant.
    }
    \label{fig:room}
\end{figure}

We consider the following scenario. The environment is an exhibition centre
where we want to reproduce the behaviour of the visitors. We consider one entry
door and one exit door. The exposition is composed of two rooms
(Fig.~\ref{fig:room}). The room 1 is accessible from the entry and the room 2
is before the exit. Visitors can stay in room 1 as long as they want before
leaving the exhibition.  As there are two rooms to visit in the exhibition
centre, we consider the transition between  these rooms. Thus, we have visitors
coming from the entry door and going to the first room. There is a transition
from the entry to  room 1 and also from room 1 to room 2.

These transitions refer to the way chaotic multispiral attractors are
analysed. As there  are multiple spirals,   and since the trajectory evolves
from one spiral to another, it is required to consider these transitions.  The
Poincaré section is composed of several components to handle these multiples
transitions. The analogy of transition between spirals is with transitions
from one room to another room to generate traces.  Moreover, this approach does not
prevent to include more components to highlight patterns as it has been done
to  provide templates of attractors   (See Fig.~15 of
\cite{rosalie2014toward}). For instance, for a Malasoma attractor bounded by a
genus one torus, a Poincar\'e section made of four components has been used to
detail torsions or permutations movements \cite{rosalie2015systematic}.  This
kind of mechanism can be used to reproduce the movements of the agents.

In this particular scenario, we consider   some parts of our model as
components of a Poincar\'e section: the entry and the exit as two components
of a Poincar\'e section and the transition between room 1 and 2 is another
component. To build the first return map, we set up the following order:  the
entry is an initial component, the transition between room 1 and 2 is a
transitional component and the exit is a final component.

\subsection{Numerical experimentation using the Lorenz system}
\label{ssec:experimental}

\begin{figure}[htb]
    \centering
    \includegraphics[width=.6\textwidth]{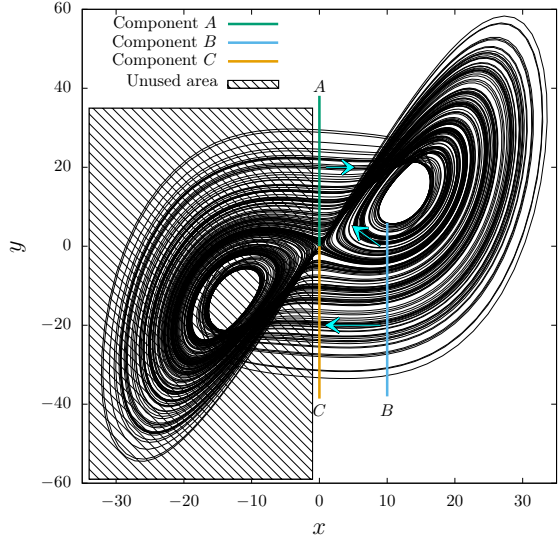}
    \caption{
        Phase portrait of an attractor solution to the Lorenz system
        (\ref{eq:lorenz}) for the parameters values $R=70$,
        $\beta=\frac{8}{3}$ and $\sigma=10$. Arrows highlight the possible
        transitions from a component to another component ($A$ to $B$, $B$ to
        $B$ or $C$).
    }
    \label{fig:lorenz_A_attra_section}
\end{figure}

We perform experimentations using the Lorenz system
\cite{lorenz1963deterministic}
\begin{equation}
  \left\{
    \begin{array}{l}
      \dot{x} = \sigma(y-x)\\
      \dot{y} = R x -y -xz \\
      \dot{z} = -\beta z + xy
    \end{array}
  \right.
  \label{eq:lorenz}
\end{equation}
in order to illustrate our methodology on a chaotic attractor bounded by a
torus with a genus higher than one.  This system is solved using a
4\textsuperscript{th} order Runge-Kutta method. We obtain an attractor
solution to this system for the parameters values $R=70$, $\beta=\frac{8}{3}$
and $\sigma=10$.  Fig.~\ref{fig:lorenz_A_attra_section} represents the
projection of the attractor in the    plane $(x, y)$. The hatch part is not
used to perform the experimentation of our method because we do not consider
the whole solution. As a consequence, a partial first return map is obtained
from the data (Fig.~\ref{fig:lorenz_A_attra_section}) with a list of three
components:
\begin{itemize}
    \item initial component $A$: Entry
    \item transitional component $B$: Transition where an agent
         decides to stay in room 1 or to proceed to room 2
    \item final component $C$: Exit
\end{itemize}

These components are given by the following equations:
\begin{equation}
    \begin{array}{l}
        \mathcal{P}_A = \{(y_n, z_n) | x_n = 0, \dot{x}_n>0 \} \\
        \mathcal{P}_B = \{(y_n, z_n) | x_n = 10, \dot{x}_n<0 \} \\
        \mathcal{P}_C = \{(y_n, z_n) | x_n = 0, \dot{x}_n<0 \}
    \end{array}
    \label{eq:components}
\end{equation}
The three components are represented Fig.~\ref{fig:lorenz_A_attra_section}
with arrows showing the flow of the attractor between them.
Here we choose arbitrarily 10 for our partition based on bounding torus theory, but
we remind that fixed point of the differential equations system can also be used to
ensure good partition according to this theory.

As  we have done for attractors bounded by high genus torus
\cite{rosalie2014toward}, we build one variable $\rho_n$ with different values
to represent the position of the component in the partial first return map
depending on its value:
\begin{itemize}
    \item initial component $A$: $\rho_n < 1$;
    \item transitional component $B$: $1 < \rho_n < 2$ and
    \item final component $C$: $2 < \rho_n < 3$.
\end{itemize}
As introduced in one of our previous work \cite{rosalie2013systematic}, we
choose to follow the orientation convention that gives the values of each
component from the inside to the outside of the attractor. This convention is
mandatory to compare chaotic mechanisms of attractors
\cite{rosalie2013systematic}. Thus, the $\rho_n$ values close to and lower than
1, 2 and 3 are respectively associated to positions in the components close to
the letters $A$, $B$ and $C$ (Fig.~\ref{fig:lorenz_A_attra_section}).

\begin{figure}[thb]
    \centering
    \includegraphics[width=.6\textwidth]{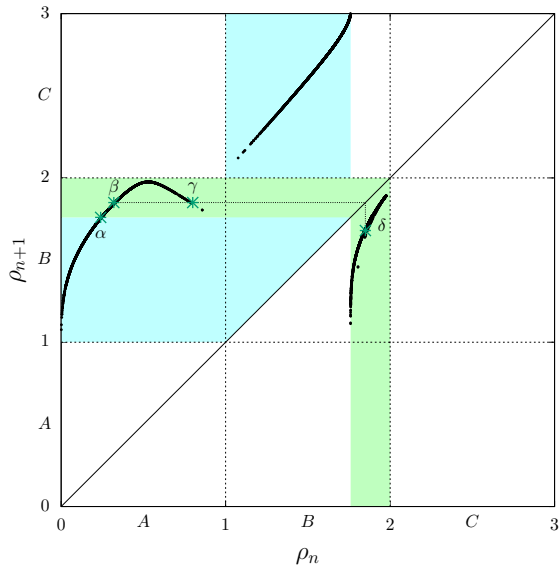}
    \caption{
        Partial first return map based on $\rho_n$ illustrating the possible
        transitions between the components. Even if some parts of this figure
        are empty, we chose to show them to make the partial first return
        map mechanism understandable. Transitions from $A$ to $B$ are possible
        and generate congestion; for instance, $\beta$ and $\gamma$ have the
        same image ($\delta$). Conversely, after the component $B$, there is
        a tearing mechanism that results in by the separation of closed
        points (the green and cyan areas underline this split).
    }
    \label{fig:lorenz_A_appli}
\end{figure}

We build  the partial first return
map (Fig.~\ref{fig:lorenz_A_appli}) based on $\rho_n$ to highlight the
possible transitions between   all
components.  It describes the transition
between components $X \to Y$ with $X$ the abscissa an $Y$ the ordinate:
\begin{itemize}
    \item  The absence of points with ordinate value for the component $A$
        underlines the fact that it is an initial component. In this map, it
        is not possible to reach component $A$.
    \item $A \to B$ in the case $(A, B)$ and $B \to B$ in the case $(B, B)$
        indicate that the next choice for the agent depends on   the value of
        component $B$: stay in room 1 or proceed to room 2.
    \item $B \to C$ in the case $(B, C)$ indicates that the agent will leave
        the room 1 to the room 2 and then leave the exhibition
    \item  The absence of points with abscissa value for the component $C$
        underlines the fact that it is a final component. From component $C$
        there is no successor point.
\end{itemize}

From the dynamical point of view, even though we do not have the whole
dynamical system, such kind of  maps gives details on the possible chaotic
mechanisms of the system. For instance, if we only consider the $(A,B)$ area,
it is an unimodal map with an increasing and a decreasing branch describing the
classical ``horseshoe'' mechanism (stretching and folding   mechanisms). This
is illustrated by the positions of the points $\beta$ and $\gamma$ in the
partial first return map (Fig.~\ref{fig:lorenz_A_appli}).  These two points
have the same ordinate but not the same abscissa. Using the map, we obtain
only one image from these points:  the point $\delta$.   In the range
$[\beta;\gamma]$ there exist pairs of points in that interval for which this
happens because there is only one image for two fibres. In terms of mobility
model analysis, it means that even if   two agents do not come from the same
position in the entry (component $A$) they can   reach the same position in
the next component $B$. Such a mechanism illustrates congestion, \textit{i.e.}
the convergence of agents to the same point.
The topological description of a chaotic attractor can be viewed as a series of mechanisms: stretching and folding are enough to generate chaos. Considering the flow of the attractor, a folding mechanism is responsible for gathering trajectories before a stretching mechanism.
The congestion is the result of a folding mechanism (continuous map with several branches) because trajectory will collapse to the same area.

Now considering points with abscissa in component $B$, their ordinate   are in
components $B$ and $C$. For an agent coming from component $B$, there is a
split into two different places. The chaotic mechanism associated to such
behaviour is the \emph{tearing mechanism}, which is well known for Lorenz
attractors (for details about mechanisms in the Lorenz system, see
\cite{byrne2004distinguishing}). The coloured areas
(Fig.~\ref{fig:lorenz_A_appli}) underline such a split directly from component
$A$ where two points  (agents) close to $\alpha$ will move  to $B$ but one
will stay in component $B$ (green area of Fig.~\ref{fig:lorenz_A_appli}) while
the other will proceed to component $C$ (cyan area of
Fig.~\ref{fig:lorenz_A_appli}). This mechanism is similar in chaotic
In the mobility model, it highlights  the
fact  that agents can stay in the same room or proceed to the last room before
leaving the exhibition.

Finally, we can use one dynamical system (\textit{e.g.} Lorenz system) to
generate the global pattern of the traces with respects to some conditions
based on the transition between areas. This is a proof of concept validated by
numerical simulations, a preliminary step before considering experimental data
instead of a solution to a dynamical system.
From data, we are expecting to find some patterns similar to tearing mechanism or folding mechanism by identify their dynamical signature in partial first return map generated from the data.
Between the mechanisms of transitions in experimental real world data we will then look for a
dynamical system with similar dynamical properties to use it as a trace generator.
The dynamical system will not be directly extracted from the data with an algorithm.
One differential equations system, with a set of parameters, has to be found in a database of research articles detailing topology of chaotic attractors (including the list of chaotic mechanisms).
The additional search of an appropriate dynamical system including equations and parameters
can be achieved using a method we developed providing templates of attractors
directly from a bifurcation diagram [40].
This article illustrates the richness of non-equivalent chaotic dynamics that could be found in one dynamical system, and it also provides various chaotic mechanisms with their parameters.
To complete, we have to mention that multiple attractors can provide same mechanisms. For instance, the multispiral attractor can be used instead of the Lorenz system because they share common chaotic mechanism.
Once a system has been found, the traces can be obtained by solving the dynamical system with
initial points (the mobile agent) in the initial components to let them evolve in the
environment by visiting the other components and escape via the final
components.
We thus propose to find a model fitting the data by giving a similar partial first return map. This model could provide a good mobility model, even if there is a suppressed or hidden part that is not used. We consider agents as particles in a flow where the chaotic dynamics provide enough variability despite the deterministic process to be used as a mobility model with our methodology.

\section{Conclusion}

In systems containing mobile entities, we emphasize the fact that chaos can be
built, as well as it can emerge, by synchronization  of the entities or  by
interaction with the environment.  In both   cases, if a chaotic  state is
observed,  then dedicated tools can be employed to analyse it. For instance,
bifurcation diagrams can illustrate transition from limit cycles (periodic
solution) to various type of chaotic dynamics.   We have seen that, Lyapunov
exponents are indicators to distinguish chaos   from hyper-chaos. However, for
chaotic dynamics, this measure is not well adapted to detail the chaotic
mechanism. Thus, others tools (Poincar\'e section or periodic orbits for
instance) leading to topological   characterizations can separate non-equivalent
chaos, and provide more accurate analysis of these chaotic
dynamics. The recent improvements concerning comparison of chaotic attractors
and topological characterization method can be used to identify or distinguish
chaotic   mechanisms and consequently identify and distinguish particular
behaviours of dynamical mobile networks.

Consequently, we   have focused on the dynamical structure exhibited by the
transitions between components of a Poincar\'e section. These transitions are
significant and reliable to design a mobility model with congestion or
distribution of agents in a given area. The procedure is applied on a Lorenz
system to highlight   the advantages of our methodology. The exhibited
structure can be used to both generate traces and analyse them.  For instance,
with the constant increase  in the number of connected devices carried by
users, there are several ways to collect real traces with their approval. It
has been done for the students of the University Politehnica of Bucharest
\cite{ciobanu2016crawdad} where their social  interactions have been studied.
In our future work we will apply our methodology to analyse agents traces
obtained by such measurements to reproduce and generate traces by finding
the most appropriate dynamical system.

\section*{Acknowledgements}

The authors also would like to thank the reviewers for their useful comments
and remarks that have made it possible to improve this article.

\printbibliography

\end{document}